\documentclass[aps,prb,twocolumn,showpacs,preprintnumbers,superscriptaddress,amsmath,amssymb]{revtex4-1}

\usepackage[utf8]{inputenc} 
\usepackage[english]{babel} 
\usepackage{graphicx}
\usepackage{amssymb,amsmath} 
\usepackage[hidelinks]{hyperref} 
\usepackage{epstopdf}
\usepackage{nicefrac}
\usepackage{dcolumn}
\usepackage{bm}
\usepackage{epsfig}
\usepackage{color}
\usepackage{siunitx} 
\usepackage{multirow}
\usepackage{physics}
\hypersetup{
	colorlinks,
	citecolor=blue,
	linkcolor=blue,
	urlcolor=blue,
}

\begin{document}

\title{Polarization resolved Cu $L_3$-edge resonant inelastic x-ray scattering of orbital and spin excitations in NdBa$_{2}$Cu$_{3}$O$_{7-\delta}$}

\author{R. Fumagalli}
\affiliation{Dipartimento di Fisica, Politecnico di Milano, Piazza Leonardo da Vinci 32, I-20133 Milano, Italy}
\author{L. Braicovich}
\affiliation{Dipartimento di Fisica, Politecnico di Milano, Piazza Leonardo da Vinci 32, I-20133 Milano, Italy}
\affiliation{ESRF, The European Synchrotron, BP 220, F-38043, Grenoble Cedex, France}
\author{M. Minola}
\affiliation{Max-Planck-Institut f\"{u}r Festk\"{o}rperforschung, Heisenbergstr. 1, 70569 Stuttgart, Germany}
\author{Y.Y. Peng}
\altaffiliation{Present address: Department of Physics and Seitz Materials Research Laboratory, University of Illinois, Urbana, IL 61801, USA}
\affiliation{Dipartimento di Fisica, Politecnico di Milano, Piazza Leonardo da Vinci 32, I-20133 Milano, Italy}
\author{K. Kummer}
\affiliation{ESRF, The European Synchrotron, BP 220, F-38043, Grenoble Cedex, France}
\author{D. Betto}
\altaffiliation{Present address: Max-Planck-Institut f\"{u}r Festk\"{o}rperforschung, Heisenbergstr. 1, 70569 Stuttgart, Germany}
\affiliation{ESRF, The European Synchrotron, BP 220, F-38043, Grenoble Cedex, France}
\author{M. Rossi}
\affiliation{Dipartimento di Fisica, Politecnico di Milano, Piazza Leonardo da Vinci 32, I-20133 Milano, Italy}
\author{E. Lefran\c{c}ois}
\affiliation{Max-Planck-Institut f\"{u}r Festk\"{o}rperforschung, Heisenbergstr. 1, 70569 Stuttgart, Germany}
\author{C. Morawe}
\affiliation{ESRF, The European Synchrotron, BP 220, F-38043, Grenoble Cedex, France}
\author{M. Salluzzo}
\affiliation{CNR-SPIN, Complesso Monte Santangelo, Via Cinthia, I-80126 Napoli, Italy}
\author{H. Suzuki}
\affiliation{Max-Planck-Institut f\"{u}r Festk\"{o}rperforschung, Heisenbergstr. 1, 70569 Stuttgart, Germany}
\author{F. Yakhou}
\affiliation{ESRF, The European Synchrotron, BP 220, F-38043, Grenoble Cedex, France}
\author{M. Le Tacon}
\affiliation{Institute of Solid State Physics (IFP), Karlsruhe Institute of Technology, D-76021 Karlsruhe, Germany}
\author{B. Keimer}
\affiliation{Max-Planck-Institut f\"{u}r Festk\"{o}rperforschung, Heisenbergstr. 1, 70569 Stuttgart, Germany}
\author{N.B. Brookes}
\affiliation{ESRF, The European Synchrotron, BP 220, F-38043, Grenoble Cedex, France}
\author{M. Moretti Sala}
\affiliation{Dipartimento di Fisica, Politecnico di Milano, Piazza Leonardo da Vinci 32, I-20133 Milano, Italy}
\author{G. Ghiringhelli}
\email{giacomo.ghiringhelli@polimi.it}
\affiliation{Dipartimento di Fisica, Politecnico di Milano, Piazza Leonardo da Vinci 32, I-20133 Milano, Italy}
\affiliation{CNR-SPIN, Dipartimento di Fisica, Politecnico di Milano, 20133 Milano, Italy}

\date{\today}

\begin{abstract}
	
	High resolution resonant inelastic x-ray scattering (RIXS) has proven particularly effective in the determination of crystal field and spin excitations in cuprates. Its strength lies in the large Cu $L_{3}$ resonance and in the fact that the scattering cross section follows quite closely the single-ion model predictions, both in the insulating parent compounds and in the superconducting doped materials. However, the spectra become increasingly broader with (hole) doping, hence resolving and assigning spectral features has proven challenging even with the highest energy resolution experimentally achievable. Here we have overcome this limitation by measuring the complete polarization dependence of the RIXS spectra as function of momentum transfer and doping in thin films of NdBa$_{2}$Cu$_{3}$O$_{7-\delta}$. Besides confirming the previous assignment of $dd$ and spin excitations (magnon, bimagnon) in the antiferromagnetic insulating parent compound, we unequivocally single out the actual spin-flip contribution at all dopings. We also demonstrate that the softening of $dd$ excitations is mainly attributed to the shift of the $xy$ peak to lower energy loss. These results provide a definitive assessment of the RIXS spectra of cuprates and demonstrate that RIXS measurements with full polarization control are practically feasible and highly informative.
	
\end{abstract}

\maketitle

\section{Introduction}

Since the discovery of copper-based high temperature superconductors (HTS)\cite{BednorzMuller1986}, huge experimental and theoretical efforts have been deployed in the quest for the understanding of microscopic mechanism of unconventional superconductivity. Despite a number of crucial findings, a conclusive and generally accepted explanation of high temperature superconductivity (SC) in cuprates is still lacking\cite{KeimerKivelsonReviewHTS}. HTS are layered materials where  superconducting CuO$_2$ planes are stacked with the so-called charge reservoirs of variable structure and composition. Undoped parent compounds are Mott insulators, with spin-$\nicefrac{1}{2}$ Cu$^{2+}$ ions forming an antiferromagnetic (AF) square lattice within the planes. These planes are, in turn, weakly coupled, so that 3D AF order sets only at temperatures much lower than the mean-field temperature set by the in-plane exchange interactions. Superconductivity arises when the number of mobile carriers (holes or electrons) is altered via chemical substitution or by varying the oxygen content in the charge reservoirs. A rich and complex phase diagram, in the doping-temperature-magnetic field ($p-T-B$) phase space results from the peculiar combination of low dimensionality and strong electronic correlations in cuprates \cite{KeimerKivelsonReviewHTS}. The interplay between antiferromagnetism and SC is intriguing: whereas magnetic fields are known to suppress SC,  AF fluctuations are considered to possibly act as glue for the Cooper pairs \cite{ScalapinoRevModPhys2012}. Consequently the full characterization of spin excitations is of paramount importance for a more conclusive explanation of HTS. 

Although optical spectroscopies and inelastic neutron scattering provided most of the experimental basis in this field for many years, more recently resonant inelastic x-ray scattering (RIXS) has brought very significant advances \cite{AmentRev2011}. In particular, thanks to remarkable technical improvements over the last two decades, high resolution RIXS revealed that short-range AF correlations persist in cuprates up to very high doping levels, across and above the superconductivity dome, despite the loss of long-range AF order \cite{LucioPRL2009,LucioMagnonRIXSPRL2010,MLTparamagnonsNatPhys,JiaNatComm}. RIXS is an orbital- and site-selective energy loss spectroscopy technique where the incident photons are tuned to a core-level absorption resonance to amplify the signal and exploit the large spin-orbit interaction of core levels. When performed at the Cu $L_3$-edge ($2p_{3/2} \to 3d$ transition) it enables momentum-dependent studies of low- and medium-energy excitations of superconducting cuprates \cite{LucioPRL2009,LucioMagnonRIXSPRL2010,ddMoretti,MLTparamagnonsNatPhys,GiacomoScience,SchlappaOrbitons2012,MinolaPRL2017}. RIXS spectra contain a variety of excitations spanning over a wide range of energies, from phonons below 100 meV \cite{Ament2011e-ph,YYPRB2015,TomPRX2016}, to magnetic excitations up to 500 meV in cuprates \cite{LucioPRL2009,LucioMagnonRIXSPRL2010},  to orbital ($dd$, crystal field) \cite{ddGiacomo2005,ddChiuzbaian2006,ddGiacomo2009,ddMoretti,SchlappaOrbitons2012,ddEllis2015,ddFatale2017,ddLee2017,Kang2019} and charge transfer \cite{Kim2004CTCuK,Ishii2005CTCuK,Wakimoto2013CTCuK,Ishii2017CTOK} excitations in the eV range. Moreover, the elastic and quasi-elastic intensities collected in RIXS spectra carry information on charge density waves or charge orders and associated excitations \cite{GiacomoScience,ChaixNatPhys2017,PengNatMat2018,MiaoPNAS2017}. 

In this framework the ERIXS spectrometer at the beam line ID32\cite{BrookesBeamLine} of the ESRF - The European Synchrotron offers a special combination of experimental capabilities: very high resolution (down to 30 meV of total instrumental bandwidth at the Cu $L_3$-edge), diffractometer-quality sample manipulation and scattering geometry flexibility, full control of the polarization of the incoming x-rays (linear horizontal, vertical and circular) and polarization analysis of the scattered photons \cite{LucioPolarimeter}. The latter capability adds extra selectivity that can be decisive in the assignment of spectral features, in particular when intrinsic broadening or quasi degeneracy of peaks makes high energy resolution partly ineffective. The polarization analysis has been applied to systems other than the cuprates, as in the case of the assignment of the energy and symmetry of $ff$ excitations in CeRh$_2$Si$_2$ \cite{Amorese2018}. 

Regarding cuprates, polarimetric measurements have confirmed the charge nature of the ordering discovered in the overdoped region of (Bi,Pb)$_2$(Sr,La)$_2$CuO$_{6+\delta}$ \cite{PengNatMat2018} and have clarified the origin of interesting RIXS signals measured in electron doped cuprates: on the one hand they confirmed the charge nature of the zone-center fast-dispersing excitations\cite{Ishii2014,WSL2014} in La$_{2-x}$Ce$_x$CuO$_4$ \cite{HeptingNature2018}, while,on the other, they assigned to spin-excitations the enhanced dynamic response at the charge order wave vector of Nd$_{2-x}$Ce$_x$CuO$_4$ \cite{SilvaNetoMM_Polarimeter}. Polarization analysis of the scattered radiation can give valuable insight into the doping evolution of various excitations: both $dd$ and spin excitations get broader upon doping and this analysis can help disentangling contributions overlapping in energy due to their intrinsic width. This has been recently done for paramagnons in Refs. \onlinecite{MinolaPRL2015,PengPRB2018}.  
Despite their relatively high energy, $dd$ excitations hold an interest for the comprehension of high $T_\textrm{c}$ superconductivity as they are related to the admixture of the 3$d$ ${3z^2-r^2}$ orbital character in the $x^2-y^2$ ground state and they can gauge the degree of two-dimensionality of the electronic structure which has empirically be related to $T_\textrm{c}$ \cite{Ohta1991,Little2007,Sakakibara2010,YYNatPhys}. 

In this manuscript we report a systematic polarization-resolved high-resolution RIXS study of low-energy spin and lattice vibrational excitations and high-energy orbital excitations in high-$T_\textrm{c}$ superconducting cuprates. We also compare the experimental findings to the calculations of the theoretical RIXS cross sections within a single-ion picture by including in the calculations the polarization dependence of all possible RIXS final states, which greatly helps the assignment of spectral features in the polarization-resolved experimental spectra. In particular, we are able to describe the different spectral contributions in terms of Stokes parameters \cite{Collett2005,Detlefs2012}.
The present manuscript is organized as follows: in Section \ref{ExperimentalMethods} we discuss the experimental details, with particular emphasis on the polarimeter. In Section \ref{Theory} we discuss the interpretation of the experimental data within the framework of the single-ion model and in terms of Stokes parameters. Finally, in Section \ref{Results} we show the experimental results and discuss separately the low-energy collective and high-energy intra-ionic excitations.

\section{Experimental methods}
\label{ExperimentalMethods}

\subsection{Samples}

The NdBa$_{2}$Cu$_{3}$O$_{7-\delta}$ (NBCO) thin films used in this work belong to the ``123'' family of high $T_\textrm{c}$ superconducting cuprates and have the same crystal strucure of YBa$_{2}$Cu$_{3}$O$_{7-\delta}$ (YBCO). Cuprates belonging to the ``123'' family crystallize in a centrosymmetric orthorombic unit cell, where CuO$_{2}$ bilayers are separated by insulating blocks composed of BaO layers and CuO chains whereas the neighboring CuO$_2$ sheets are separated by a Nd ion each. Because the lattice parameters $a$ and $b$ are nearly identical in epitaxial films grown on tetragonal substrates (3.9 \AA) and $c$ = 11.7 \AA, we can neglect the orthorhombicity and adopt a tetragonal description. Similarly to YBCO, by altering the oxygen content in the CuO chains and the excess of Nd at the Ba sites (Nd$_{1+x}$Ba$_{2-x}$Cu$_3$O$_7$) it is possible to change the carrier density (holes) in the CuO$_2$ planes and obtain superconductivity with $T_\textrm{c}$ up to 95 K at optimal doping. In this work we have measured undoped (AF), underdoped (UD, $T_\textrm{c}$ =  63 K and hole concentration $p$ = 0.11) and optimally doped (OP, $T_\textrm{c}$ =  90 K and $p$ = 0.17) NBCO epitaxial films deposited by high oxygen pressure diode sputtering on a (001) SrTiO$_3$ single crystal substrate with an almost perfect in plane matching of the lattice parameters. More details on samples growth and characterization have already been given in Refs. \onlinecite{Salluzzo1998NBCO} and \onlinecite{Salluzzo2005NBCO}.

\subsection{RIXS measurements}

\begin{figure}[ht]
	\includegraphics[width=1\columnwidth]{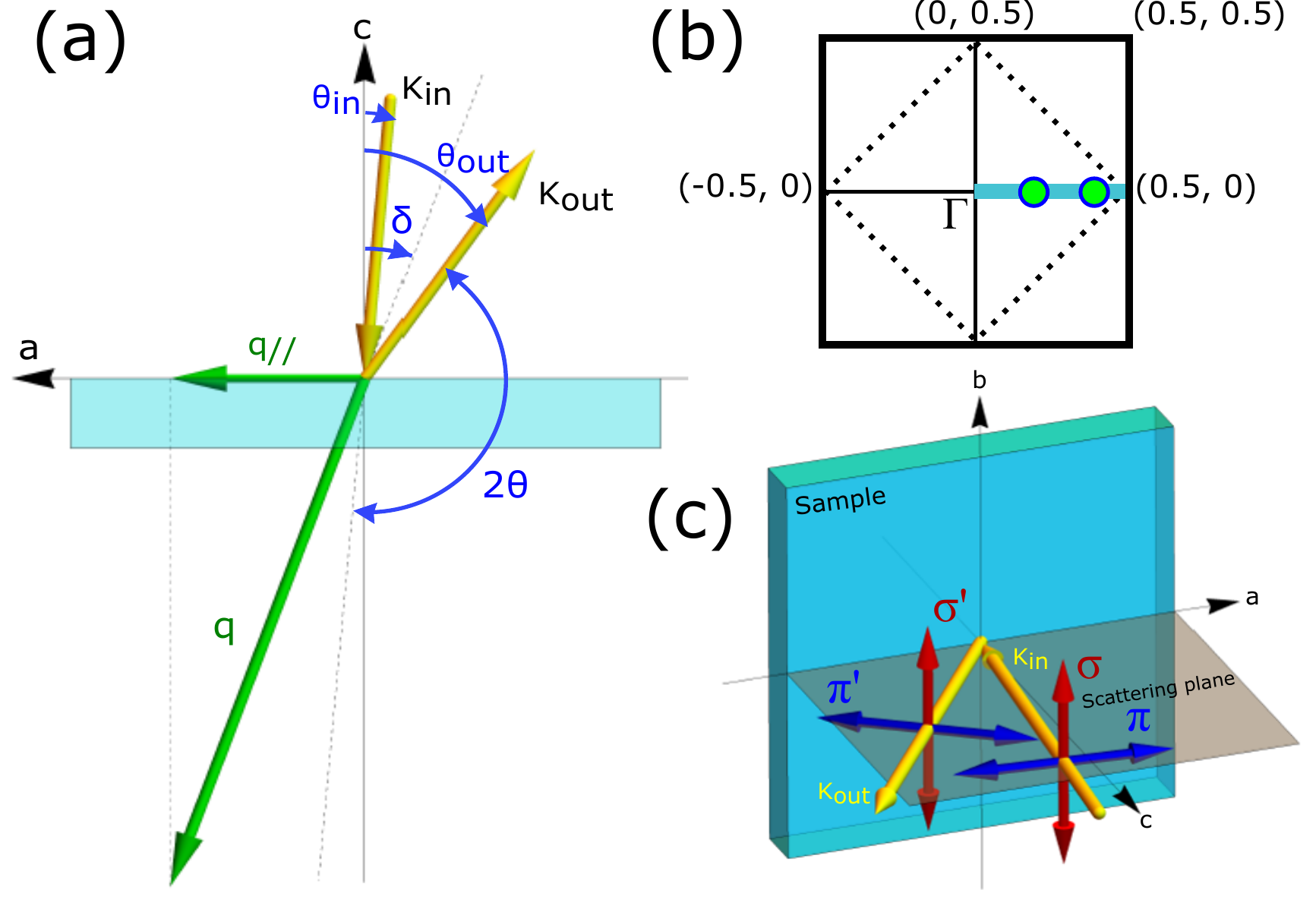}
	\caption{\label{fig:ScatteringGeometry} (a) Sketch of the scattering geometry. (b) Schematic representation of the first Brillouin zone (the solid line refers to the crystal lattice, dots refer to the magnetic lattice). The two circles along the [1 0 0] direction represent the in-plane momentum values of experimental data. (c) Representation of the experimental geometry which highlights the possibility to perform polarization analysis of the scattered light.}
\end{figure}

RIXS spectra have been acquired with an overall energy resolution of $\sim$ 80 meV, as determined by measuring the non-resonant response of silver paint placed on a corner of the sample surface. The incident photon energy was tuned to the Cu $L_3$-edge ($\sim$ 931 eV) and the polarization of the radiation could be set either parallel ($\pi$, horizontal) or perpendicular ($\sigma$, vertical) to the scattering plane. The scattering geometry is sketched in Fig.\,\ref{fig:ScatteringGeometry}(a), where the scattering angle ($2\theta$) was fixed at 149.5$^\circ$ in order to maximize the momentum transfer. Note that the momentum transfer is expressed in terms of its projection onto the CuO$_2$ plane ($q_{\|}$), the relevant quantity in quasi-two dimensional (2D) cuprates. The scattering occurs in the sample $ac$ plane, such that $q_{\|}$ could be changed by simply rotating the sample around its $b$-axis; if $\delta$ is the angle between the sample $c$-axis and the total momentum transfer $\textbf{q}$, then $q_{\|}=q \sin\delta$. The values of transferred momentum are expressed in reciprocal lattice units (rlu), defined in units of reciprocal lattice vectors 2$\pi/a$, 2$\pi/b$ 2$\pi/c$, where $a$, $b$ and $c$ are the lattice constants of the unit cell of NBCO. The experimental data shown in this work were collected at $q_{\|}$ = 0.2 rlu and 0.4 rlu along the anti-nodal $\Gamma \longrightarrow X$ or [1 0 0] direction, \emph{e.g.} parallel to the Cu-O bonds in the CuO$_2$ planes, as depicted in panel (b) of Fig.\,\ref{fig:ScatteringGeometry}. All measurements were collected at 20 K.

Throughout our paper, we will present RIXS intensities (corrected for self-absorption as explained in Section \ref{self_absorption}) in units of eV$^{-1}$srad$^{-1}$, \emph{i.e.} by normalizing the spectra to the collection solid angle and incident photon flux, and by taking into account the sampling frequency (energy, in our case) and the spectrometer efficiency. The collection solid angle of ERIXS (5$\cdot 10^{-5}$ srad) is given by the product of the angular acceptance of the grating ($\sim 2.5$ mrad) and of the collimating mirror (20 mrad). The photon flux was estimated by measuring and calibrating the drain current generated by the beam in the last optical element before the sample; it amounts to approximately $ 10^{12}$ photons/s in a bandwidth of 45 meV. The sampling frequency of the RIXS spectra is $10$ meV, as determined by the pixel pitch of the detector and the dispersion rate of the grating. Finally, the spectrometer efficiency is limited by the reflectivity of the grating and amounts to approximately 0.1 for the 1400 mm$^{-1}$ grating \cite{BrookesBeamLine} used in the present work. We believe that presenting RIXS intensities in terms of ``scattering probability'' will facilitate the comparison of data taken with different experimental setups (different synchrotron sources, beamlines, spectrometers, etc.).

\subsection{Polarimeter}

The advantages of performing polarization resolved measurements have been demonstrated with the prototype of a polarization-selective optical element (hereafter referred to as polarimeter) previously installed on the AXES spectrometer at the ID08 beamline of the ESRF\cite{LucioPolarimeter,MinolaPRL2015}. Similarly, the new polarimeter installed on the ERIXS spectrometer at ID32 is based on a W/B$_4$C multilayer mirror. Unlike AXES, however, ERIXS delivers horizontally collimated radiation, \emph{e.g.} the incidence angle on the multilayer is the same for all photons. We therefore adopted a graded multilayer, which period changes linearly along one direction in its surface (consistently with the photon energy dispersion of the spectrometer at the position of the multilayer), keeping the reflectivity constant over several eVs. The nominal working angle of the multilayer mirror is $\sim$20$^\circ$, corresponding to reflectivities $r_{\pi^\prime}$ = 0.085 and $r_{\sigma^\prime}$ = 0.140 for the $\pi$ and $\sigma$ components of the polarization relative to the multilayer scattering plane. The average efficiency of the device is, therefore, $r_0 = (r_{\sigma^\prime}+r_{\pi^\prime})/2 \approx 11.2\%$. Note that the full suppression of one polarization component ($r_{\pi^\prime}$ = 0) could be obtained at the Brewster angle (45$^\circ$). However, the reflectivity of the unsuppressed polarization component is also very much reduced to $r_{\sigma^\prime}$ = 0.012, too low for practical usage of the polarimeter. Still, a consistent decomposition of RIXS spectra in terms of outgoing photon polarization components can be done, as explained below. In the present case, polarization-resolved RIXS spectra with good statistics could be acquired in 180 minutes of accumulation time, while polarization-unresolved RIXS spectra were counted for 30 minutes.

In panel (c) of Fig.\,\ref{fig:ScatteringGeometry} we show the linear components of the incident and scattered photon polarization, giving rise to four distinct scattering geometries: $\pi\pi'$, $\pi\sigma'$, $\sigma\sigma'$ and $\sigma\pi'$, where the unprimed (primed) symbols refer to the incident (scattered) photon polarization. In the following, we will refer to $\sigma\pi'$ and $\pi\sigma'$ as the cross-polarization channels and to  $\sigma\sigma'$ and $\pi\pi'$ as the non-cross-polarization channels.

\subsection{Polarization analysis and Poincaré-Stokes parameters}\label{polanalysis_procedure}

The polarization state of the electromagnetic radiation scattered by the sample in the RIXS process is described by the (complex) components $E_{\sigma^\prime}=|E_{\sigma^\prime}|e^{\imath \delta_{\sigma^\prime}}$ and $E_{\pi^\prime}=|E_{\pi^\prime}|e^{\imath \delta_{\pi^\prime}}$ of the electric field in the coordinate axes $\bm{\epsilon}_{\sigma^\prime}$ and $\bm{\epsilon}_{\pi^\prime}$, perpendicular to the propagation direction $\mathbf{k}^\prime$. Unfortunately, these are not directly accessible, because the phase information is lost when measuring intensities. Instead, it is often convenient to adopt a description of the polarization based on the so-called Stokes parameters (see Appendix \ref{PoincareStokes}), which are measurable quantities\cite{Collett2005,Detlefs2012}. 

As already mentioned earlier, the polarization analysis is achieved by exploiting the difference in the reflectivity of a multilayer mirror for the $\sigma^\prime$ and $\pi^\prime$ components of the radiation. However, the multilayer does not completely suppress any of the two photon polarizations, but rather attenuates one more than the other ($r_{\pi^\prime} \approx r_{\sigma^\prime}/2$). A polarization-resolved RIXS spectrum is, therefore, obtained by combining two independent measurements of the intensity of the polarization-unresolved or ``direct'' beam  ($I$) and the intensity of the beam after being reflected by the multilayer mirror ($I_M$). We introduce the Stokes vector for the direct beam
\begin{equation}\label{StokesD}
\mathbf{S} = \begin{bmatrix}
S_{0}\\
S_{1}\\
S_{2}\\
S_{3}
\end{bmatrix} = \begin{bmatrix}
|E_{\sigma^\prime}|^2+|E_{\pi^\prime}|^2\\
|E_{\sigma^\prime}|^2-|E_{\pi^\prime}|^2\\
\frac{1}{2}\left(|E_{\sigma^\prime}+E_{\pi^\prime}|^2-|E_{\sigma^\prime}-E_{\pi^\prime}|^2 \right)\\
\frac{1}{2}\left(|E_{\sigma^\prime}-\imath E_{\pi^\prime}|^2-|E_{\sigma^\prime}+\imath E_{\pi^\prime}|^2 \right)
\end{bmatrix}
\end{equation}
where $S_{0}=|E_{\sigma^\prime}|^2+|E_{\pi^\prime}|^2=I$ is the total intensity of the scattered radiation as measured on the detector. In addition, one can define the Poincaré-Stokes parameters as $P_{i}=S_{i}/S_{0}$, with $i=1,2,3$. In case of fully polarized radiation, as in our case, $S_{1}^2 + S_{2}^2 + S_{3}^2 = S_{0}^2$ ($P_{1}^2 + P_{2}^2 + P_{3}^2 = 1$) and the Poincaré-Stokes parameters define a point on the Poincaré sphere of unit radius (see Fig.\,\ref{fig:Ellipse} in Appendix \ref{PoincareStokes}). The Stokes parameters of the beam reflected by the multilayer can be calculated from $\mathbf{S}_M=\mathcal{M} \mathbf{S}$, where $\mathcal{M}$ is the M\"{u}ller matrix for reflection by an optical elements (see Appendix \ref{PoincareStokes}). We obtain
\begin{equation}\label{StokesM}
\mathbf{S}_M = \begin{bmatrix}
S_{M,0}\\
S_{M,1}\\
S_{M,2}\\
S_{M,3}
\end{bmatrix} =\begin{bmatrix}
\frac{1}{2}\left[r_{\sigma^\prime}(S_{0}+S_{1}) + r_{\pi^\prime}(S_{0}-S_{1})\right] \\
\frac{1}{2}\left[r_{\sigma^\prime}(S_{0}+S_{1}) - r_{\pi^\prime}(S_{0}-S_{1})\right] \\
\sqrt{r_{\sigma^\prime} r_{\pi^\prime}}S_{2}\\
\sqrt{r_{\sigma^\prime} r_{\pi^\prime}}S_{3}
\end{bmatrix},
\end{equation}
where $S_{M,0}=I_M$ is the total intensity of the beam after being reflected by the multilayer mirror as measured on the detector. It is important to point out that, in general, all polarization states contribute to $I_M$ via their projections on the $\bm{\epsilon}_{\sigma^\prime}$ and $\bm{\epsilon}_{\pi^\prime}$ coordinate axes. In the following, we assume that the scattered radiation is fully linearly polarized, implying that $|S_{1}|=S_{0}$ ($|P_{1}|=1$) and $S_{2}=S_{3}=0$ ($P_{2}=P_{3}=0$). In Section \ref{singleion}, we will show that this assumption is well justified in all, but a very few cases for Cu$^{2+}$ RIXS cross-sections in the adopted scattering geometry. Under this assumption,
\begin{equation}\label{Isigmapi}
I_{\sigma^{\prime},\pi^{\prime}}= \frac{S_{0}\pm S_{1}}{2}= \frac{I\pm S_{1}}{2},
\end{equation}
with
\begin{equation}\label{S1prime}
S_{1}=\frac{1}{A}\left(\frac{S_{M,0}}{r_0}-S_{0}\right)= \frac{1}{A}\left(\frac{I_M}{r_0}-I\right)
\end{equation}
and 
\begin{equation}
A = \frac{r_{\sigma^\prime}-r_{\pi^\prime}}{r_{\sigma^\prime}+r_{\pi^\prime}} = \frac{r_{\sigma^\prime}-r_{\pi^\prime}}{2r_0}
\end{equation}
is the multilayer polarization sensitivity ($A\approx 0.25$), equivalent to the Sherman function of Mott detectors for spin-resolved photoemission experiments \cite{Kessler1985PolarizedElectrons}. 

\begin{figure}
	\includegraphics[width=1\columnwidth]{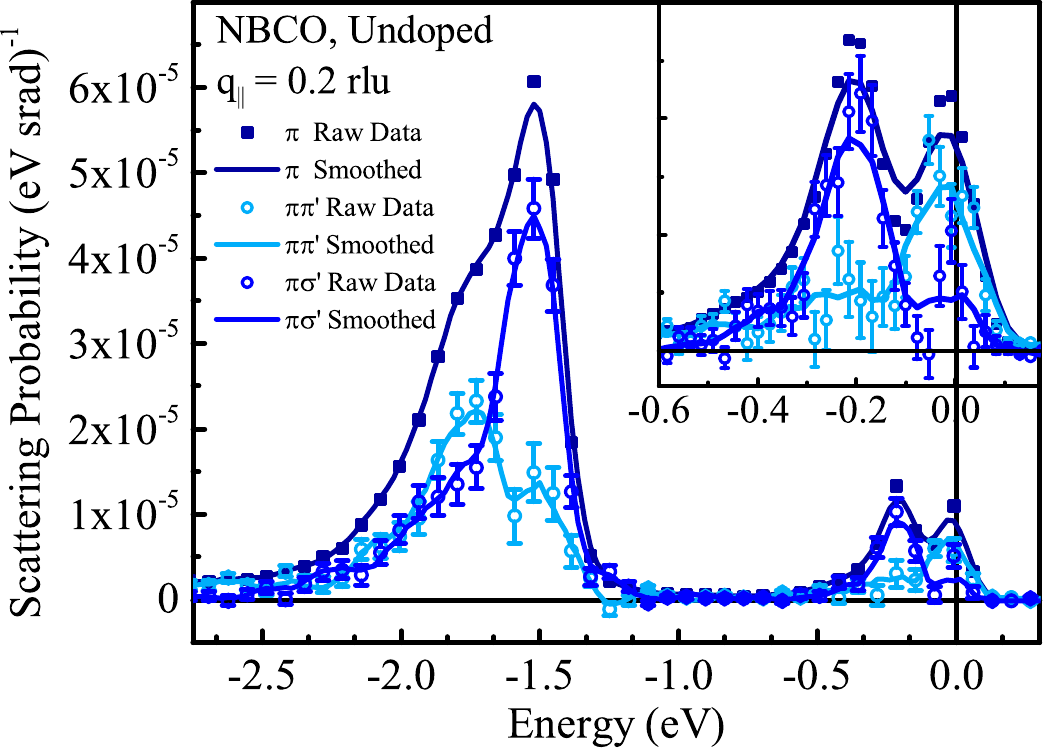}
	\caption{\label{fig:errorbars} Example of polarization resolved RIXS spectrum of AF NBCO measured with $\pi$ polarization at in-plane momentum equal to 0.2 rlu. For clarity, we show only every second data point (symbols). Inset: the close-view of the low energy region is shown. Error bars are calculated on raw data (symbols), while continuous lines represent data smoothed on 7 points.}
\end{figure}

In Fig.\,\ref{fig:errorbars} we show the polarization-resolved RIXS spectra of AF NBCO at $q_{\|}$ = 0.2 rlu for $\pi$ incident photon polarization (error bars are calculated as explained in appendix \ref{Errorbars}). The RIXS spectra are dominated by intense features in the energy loss window between -2.5 and -1.0 eV, which are usually ascribed to crystal field excitations\cite{}. In the low-energy region, other excitations are also clearly visible and highlighted in the inset. One can distinguish a (quasi-)elastic line at zero energy loss and magnetic excitations below 0.5 eV. A quick inspection of the polarization-resolved RIXS spectra immediately shows that different excitations have their own polarization dependences. In particular, we note that magnetic excitations occur mainly in the $\pi\sigma^\prime$ polarization channel, while the (quasi)elastic line is observed in the non-cross polarization channel. Crystal-field excitations also have a clear polarization dependence, which will be exploited in the following when discussing the case of doped cuprates.

\subsection{Self-absorption corrections}\label{self_absorption}

The decomposition of the RIXS spectra in terms of outgoing photon polarization components offers the possibility to correct self-absorption effects in a more reliable way than in unpolarized RIXS. Indeed, self-absorption alters the shape of the spectrum because photons are more strongly re-absorbed at small energy losses than at large ones due to the resonance. Since the absorption coefficient of a material depends on both the photon energy and polarization, so does the self-absorption. The knowledge of the scattered photon polarization is, therefore, an important ingredient for accurate self-absorption corrections.

We here take full advantage of the polarization resolution and apply self-absorption corrections to all experimental RIXS spectra shown in this paper. Following the procedure discussed in detail in the Supplemental Material of Ref.\,\onlinecite{MinolaPRL2015}, we define a correction factor $C_{\bm{\epsilon},\bm{\epsilon}^\prime}(\omega_1,\omega_2)$, which depends on the energy $\omega_1$ ($\omega_2$) and polarization $\bm{\epsilon}$ ($\bm{\epsilon}^\prime$) of the incident (scattered) photons, such that the corrected RIXS intensity $I_{\bm{\epsilon}^\prime}^\mathrm{corr}(\omega_2)$ is related to the measured intensity $I_{\bm{\epsilon}^\prime}^\mathrm{meas}(\omega_2)$ by
\begin{equation}
I_{\bm{\epsilon}^\prime}^\mathrm{corr}(\omega_2) = \frac{I_{\bm{\epsilon}^\prime}^\mathrm{meas}(\omega_2)}{C_{\bm{\epsilon},\bm{\epsilon}^\prime}(\omega_1,\omega_2)}.
\end{equation}
In particular, the correction factor is given by
\begin{equation}
C_{\bm{\epsilon},\bm{\epsilon}^\prime}(\omega_1,\omega_2) =\frac{1}{1+t_{\bm{\epsilon},\bm{\epsilon}^\prime}(\omega_1,\omega_2)u},
\label{equ:C}
\end{equation}
where $u=\cos(\theta_{in})/\cos(\chi)$ is a geometrical factor depending on the photon angles of incidence ($\theta_{in}$) and scattering ($\chi$) as measured from the normal to the sample surface (see Fig.\,\ref{fig:ScatteringGeometry}(a)), and 
\begin{equation}
t_{\bm{\epsilon},\bm{\epsilon}^\prime}(\omega_1,\omega_2) =\frac{\alpha_0+\alpha_{\bm{\epsilon}^\prime}(\omega_2)}{\alpha_0+\alpha_{\bm{\epsilon}}(\omega_1)}
\label{equ:ru}
\end{equation}
where $\alpha_0$ and $\alpha_{\bm{\epsilon}}(\omega)$ are the non-resonant and resonant part of the absorption coefficient, respectively, which can be experimentally determined. Given the large orbital anisotropy of the cuprates, the absorption coefficient $\alpha_{\bm{\epsilon}}(\omega)$ varies enormously depending on the orientation of $\bm{\epsilon}$ with respect to the sample crystallographic directions. At the $L_3$-edge resonance, the absorption is minimized (maximized) for $\bm{\epsilon}\|c$ ($\bm{\epsilon}\perp c$)\cite{Chen1992}. For this reason, the knowledge of the polarization of the scattered photons allows one to determine the self-absorption correction coefficient in a precise way.

\begin{figure}[ht]
	\includegraphics[width=1\columnwidth]{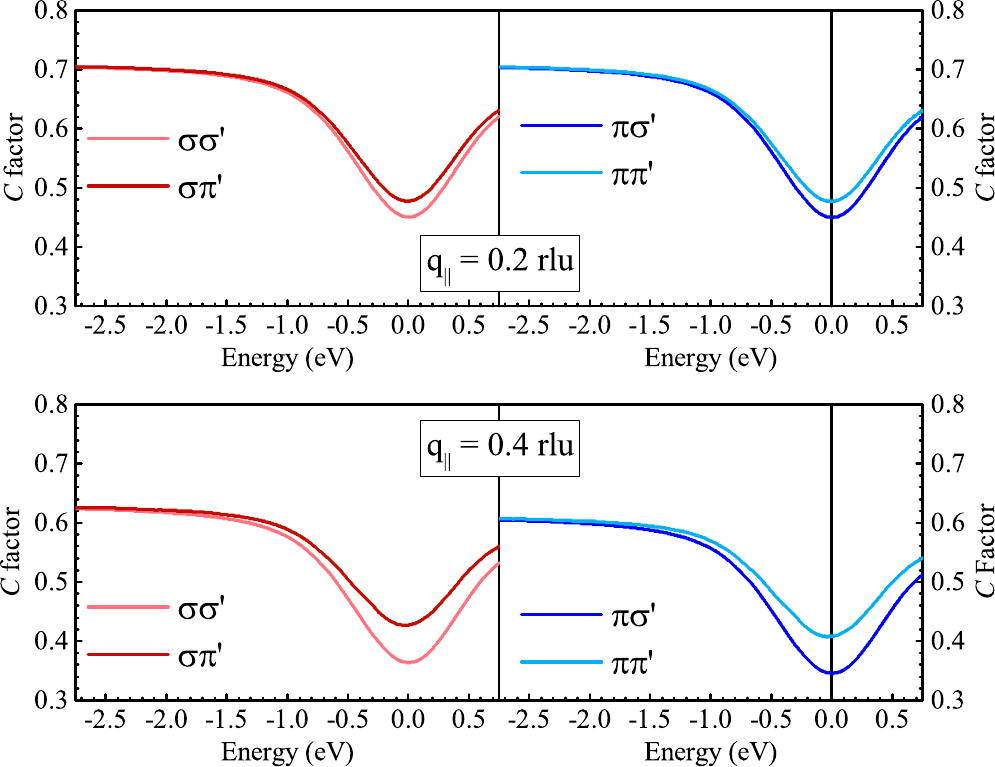}
	\caption{\label{fig:Cfactor} Self-absorption correction factors at $q_{\|}=0.2$ rlu (top panels) and at $q_{\|}=0.4$ rlu (bottom panels) for both $\sigma$ and $\pi$ polarization of the incident and scattered light calculated from the XAS spectra measured by total electron yield on the AF NBCO sample.}
\end{figure}

We show in Fig.\,\ref{fig:Cfactor} the correction factors at $q_{\|}=0.2$ rlu and at $q_{\|}=0.4$ rlu for both $\sigma$ and $\pi$ polarizations of the light. Clearly, the self-absorption correction affects mostly the low energy spectral range, leaving the $dd$ lineshape unchanged. For all the cases, the difference between the two possible polarization states of the scattered light does not exceed $\approx 10-20\%$.

\section{Theory}
\label{Theory}

\subsection{Single-ion model for the calculation of the RIXS cross-sections}\label{singleion}

To help the interpretation of polarization-resolved RIXS spectra, we adopt a single-ion model that allow us to calculate $L_3$-edge RIXS cross-sections. The case of cuprates is particularly straightforward, because a Cu$^{2+}$ (3$d^9$) ion in $D_{4h}$ symmetry has initial, intermediate and final states with one hole filling the ${x^2-y^2}$ orbital, one of the four spin-orbit coupled 2$p_{3/2}$ orbitals and one of the five crystal-field-split 3$d$ orbitals, respectively. Therefore, the Kramers-Heisenberg formula for the calculation of the RIXS amplitudes $\mathcal{A}_{\bm{\epsilon}^\prime}$ of the various final states can be easily implemented.

The use of the Cu$^{2+}$ single-ion model has already been proved quite powerful. It predicted the possibility to measure single spin-flip excitations due to the strong spin-orbit coupling of the 2$p_{3/2}$ core hole \cite{Ament2009,HaverkortTheoryMagnon2010} and found immediate experimental confirmation \cite{LucioMagnonRIXSPRL2010,LucioPRB2010}. The single-ion model was also useful to assign crystal-field excitations in RIXS data of cuprates based on their dependence on incident photon polarization and scattering geometry; a detailed discussion can be found, for example, in Ref. \onlinecite{ddMoretti}. Here, we use the explicit dependence of the RIXS cross sections of the scattered photon polarization to calculate the Stokes vector
\begin{equation}
\mathbf{S}=\begin{bmatrix}
S_{0}\\
S_{1}\\
S_{2}\\
S_{3}
\end{bmatrix} \propto \begin{bmatrix}
 |\mathcal{A}_{\sigma^\prime}|^2+|\mathcal{A}_{\pi^\prime}|^2\\
|\mathcal{A}_{\sigma^\prime}|^2-|\mathcal{A}_{\pi^\prime}|^2\\
|\mathcal{A}_{\sigma^\prime}\!+\!\mathcal{A}_{\pi^\prime}|^2-|\mathcal{A}_{\sigma^\prime}\!-\!\mathcal{A}_{\pi^\prime}|^2\\
|\mathcal{A}_{\sigma^\prime}\!-\!\imath\mathcal{A}_{\pi^\prime}|^2-|\mathcal{A}_{\sigma^\prime}\!+\!\imath \mathcal{A}_{\pi^\prime}|^2
\end{bmatrix},
\end{equation}
and the corresponding Poincaré-Stokes parameters $\mathbf{P}$ ($P^\prime_i = S_i/S_0$, $i=1,2,3$). The latter are reported in Tables \ref{table:StokesSigma} and \ref{table:StokesPi} for the scattering geometry depicted in Fig.\ \ref{fig:ScatteringGeometry}(a) and (c) for $\pi$ and $\sigma$ incident photon polarization, respectively. The ground state is assumed to be $x^2-y^2 \downarrow$, such that $x^2-y^2 \downarrow$ indicates elastic scattering (the arrow direction indicates the spin state). Excited states may involve a spin flip (\emph{e.g.}, $x^2-y^2 \uparrow$), an orbital change (\emph{e.g.}, $3z^2-r^2 \downarrow$) or both (\emph{e.g.}, $3z^2-r^2 \uparrow$). 

\begin{table}[ht]
	\begin{tabular}{lccc}
		\hline\hline
		&$P_{1}$ &$P_{2}$&$P_{3}$ \\ \hline 
		$x^2-y^2 \downarrow$&1 & 0 & 0\\
		$x^2-y^2 \uparrow$&-1 & 0 & 0\\
		$xy\downarrow$&-1 & 0 & 0\\
		$xy\uparrow$&1 & 0 & 0\\
		$xz\downarrow$&-1 & 0 & 0\\
		$xz\uparrow$&-1 & 0 & 0\\
		$yz\downarrow$&$\frac{4\cos(2\theta_o)-3}{1+8\sin^2\theta_o}$ & 0 & $\frac{4\sqrt{2}\sin\theta_o}{1+8\sin^2\theta_o}$\\
		$yz\uparrow$&1 & 0 & 0\\
		$3z^2-r^2\downarrow$&$\frac{1}{2}\frac{\cos(2\theta_o)+3}{2+\sin^2\theta_o}$& 0 & $\frac{2\sqrt{2}\sin\theta_o}{2+\sin^2\theta_o}$ \\
		$3z^2-r^2\uparrow$&-1 & 0 & 0\\
		\hline\hline
	\end{tabular}
	\caption{Poincaré-Stokes parameters of scattered radiation calculated within the Cu$^{2+}$ single-ion model with $\sigma$ incident photon polarization for the various final states. $\theta_o$ is the angle between the scattered photons and the normal to the sample surface.}
	\label{table:StokesSigma}
\end{table}

\begin{table}[ht]
	\begin{tabular}{lccc}
		\hline\hline
		&$P_{1}$ &$P_{2}$&$P_{3}$ \\ \hline 
		$x^2-y^2 \downarrow$&-1 & 0 & 0 \\
		$x^2-y^2 \uparrow$&1 & 0 & 0\\
		$xy\downarrow$&1 & 0 & 0\\
		$xy\uparrow$&-1 & 0 & 0\\
		$xz\downarrow$&-1 & 0 & 0\\
		$xz\uparrow$&-1 & 0 & 0\\
		$yz\downarrow$&1 & 0 & 0\\
		$yz\uparrow$&$\frac{1-2\sin^2\theta_o}{1+2\sin^2\theta_o}$  & 0 & $\frac{2\sqrt{2}\sin\theta_o}{1+2\sin^2\theta_o}$\\
		$3z^2-r^2\downarrow$&-1 & 0 & 0\\
		$3z^2-r^2\uparrow$&$\frac{1-2\sin^2\theta_o}{1+2\sin^2\theta_o}$ & 0 & $\frac{2\sqrt{2}\sin\theta_o}{1+2\sin^2\theta_o}$ \\
		\hline\hline
	\end{tabular}
	\caption{Poincaré-Stokes parameters of scattered radiation calculated within the Cu$^{2+}$ single-ion model with $\pi$ incident photon polarization for the various final states. $\theta_o$ is the angle between the scattered photons and the normal to the sample surface.}
	\label{table:StokesPi}
\end{table}

It is interesting to note that, when the scattering occurs in the sample $ac$ ($xz$) plane with full $\sigma$ or $\pi$ incident photon polarization, the scattered radiation is also fully $\sigma^\prime$ or $\pi^\prime$ polarized ($P_{1} = 1$ or -1, respectively, and $P_{2}=P_{3}=0$) for all excited states, but two: $yz,\downarrow$ ($yz,\uparrow$) and $3z^2-r^2,\downarrow$ ($3z^2-r^2,\uparrow$) for $\sigma$ ($\pi$) incident photon polarization. Here, $S_{3} \neq 0$ ($P_{3} \neq 0$) and the polarization analysis described in Section \ref{polanalysis_procedure} might be ambiguous. For example, if we assume that the scattered radiation is fully circularly polarized, then $|S_{3}|=S_{0}$ ($|P_{3}|=1$) and $S_{1}=S_{2}=0$ ($P_{1}=P_{2}=0$): the intensity recorded on the detector after reflection from the multilayer mirror will be $S_{M,0}=r_0 S_{0}$. Eqs.\,(\ref{Isigmapi}) and (\ref{S1prime}) will provide $I_{\sigma^\prime}=I_{\pi^\prime}=S_{0}/2$, \emph{i.e.} the scattered intensity is equally distributed between the two polarization channels.

\begin{figure}[ht]
	\includegraphics[width=1\columnwidth]{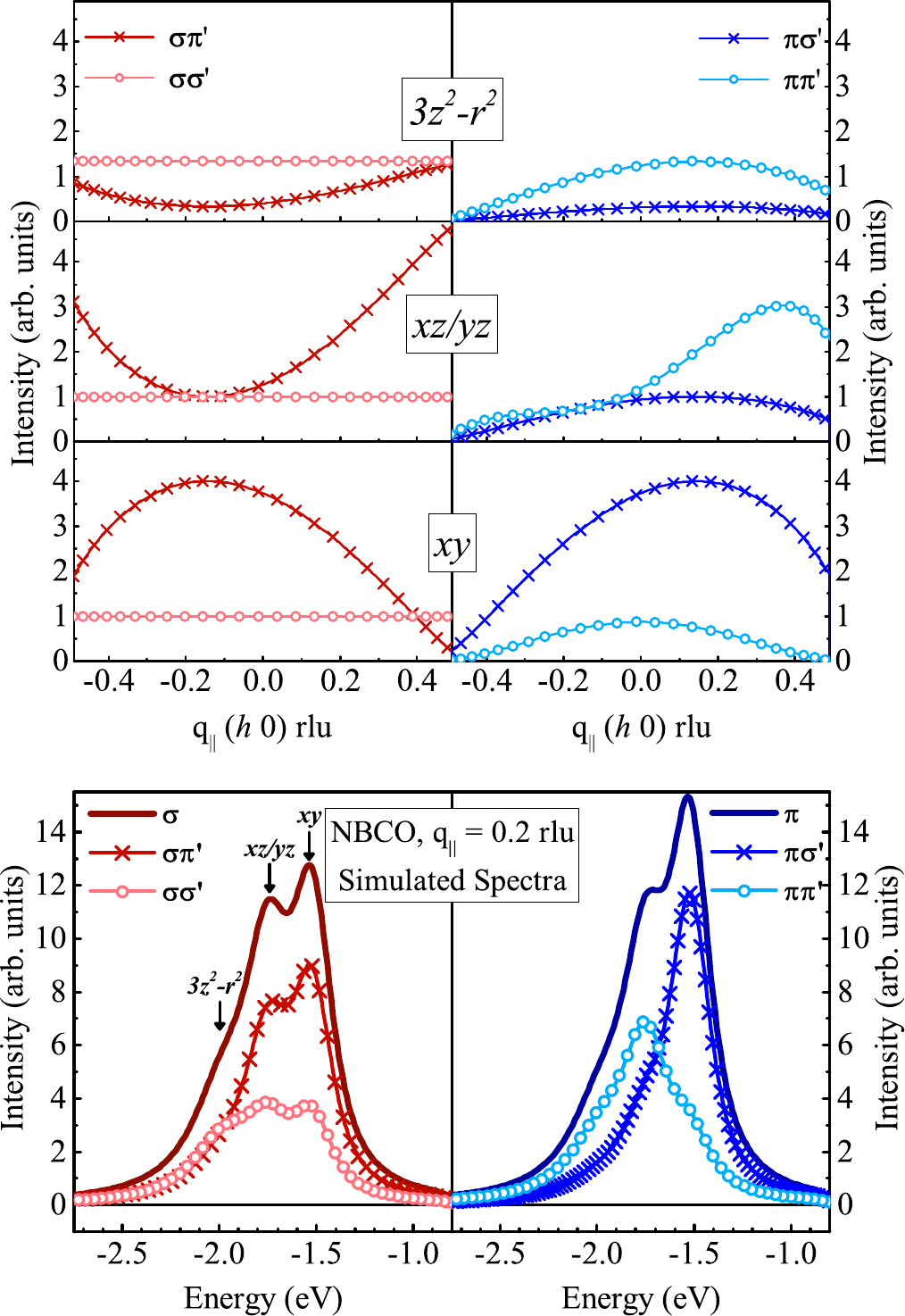}
	\caption{\label{fig:Theory} The top panel shows the in plane momentum dependence of Cu $L_{3}$-edge polarized RIXS cross sections within single-ion model for $dd$ excitations (${xy}$, doubly degenerate ${xz/yz}$ and $3z^2-r^2$ orbitals). For each curve we consider as degenerate the final states with $\Delta S = 0$ (spin-conserving) and $\Delta S = 1$ (spin-flip). For all cases the scattering angle (2$\theta$) has been fixed to 149.5$^\circ$. In the bottom panels, simulated spectra with outgoing polarization analysis of AF NBCO are sketched at $q_{\|}$ = 0.2 rlu along [1 0 0] direction of the Brillouin zone with both $\sigma$ (left) and $\pi$ (right) incident photon polarization of the light.}
\end{figure}

In the upper panel of Fig.\,\ref{fig:Theory} we show the calculated RIXS cross-sections for various final states with polarization resolution as a function of $q_{\|}$ along the [1 0 0] direction at a fixed scattering angle of $149.5^\circ$. These are summed over spin-flip and non-spin-slip states, which eventually leads us to consider only three excited states, \emph{e.g.} the ${xy}$, the ${xz/yz}$ and the ${3z^2-r^2}$ states. In the bottom panel, we simulate the RIXS spectra of AF NBCO at $q_{\|}$ = 0.2 rlu by taking the energy position and the Lorentzian lifetime broadening of the three excited state from the literature: in particular, the transition to the $xy$ state is found at -1.52 eV, the $xz/yz$ state at -1.75 eV and the ${3z^2-r^2}$ state at -1.98 eV \cite{ddMoretti}. In order to facilitate the comparison with the experiment (Fig.\,\ref{fig:errorbars}), the simulated RIXS spectra are convoluted with a Gaussian function with full width at half maximum of 80 meV that takes into account the finite experimental energy resolution. The agreement between measured and calculated RIXS spectra is remarkable, including the cases with polarization resolution of the scattered photons. The main experimental features are well reproduced, in particular the fact that with $\pi$ incident photon polarization the transitions to the $xz/yz$ and the ${3z^2-r^2}$ excited states mainly occur in the $\pi\pi^\prime$ polarization channel, while the transition to the $xy$ excited state occurs mainly belongs to the $\pi\sigma^\prime$ polarization-channel. 

The agreement between measurements and calculations shows that the single-ion model captures the symmetry of the ground and excited states of the system and provides a good description of the RIXS process. Therefore, we will use it extensively in the following to discuss in detail the photon polarization and doping dependence of spectral features in NBCO.

\begin{figure*}[ht]
	\includegraphics[width=1.5\columnwidth]{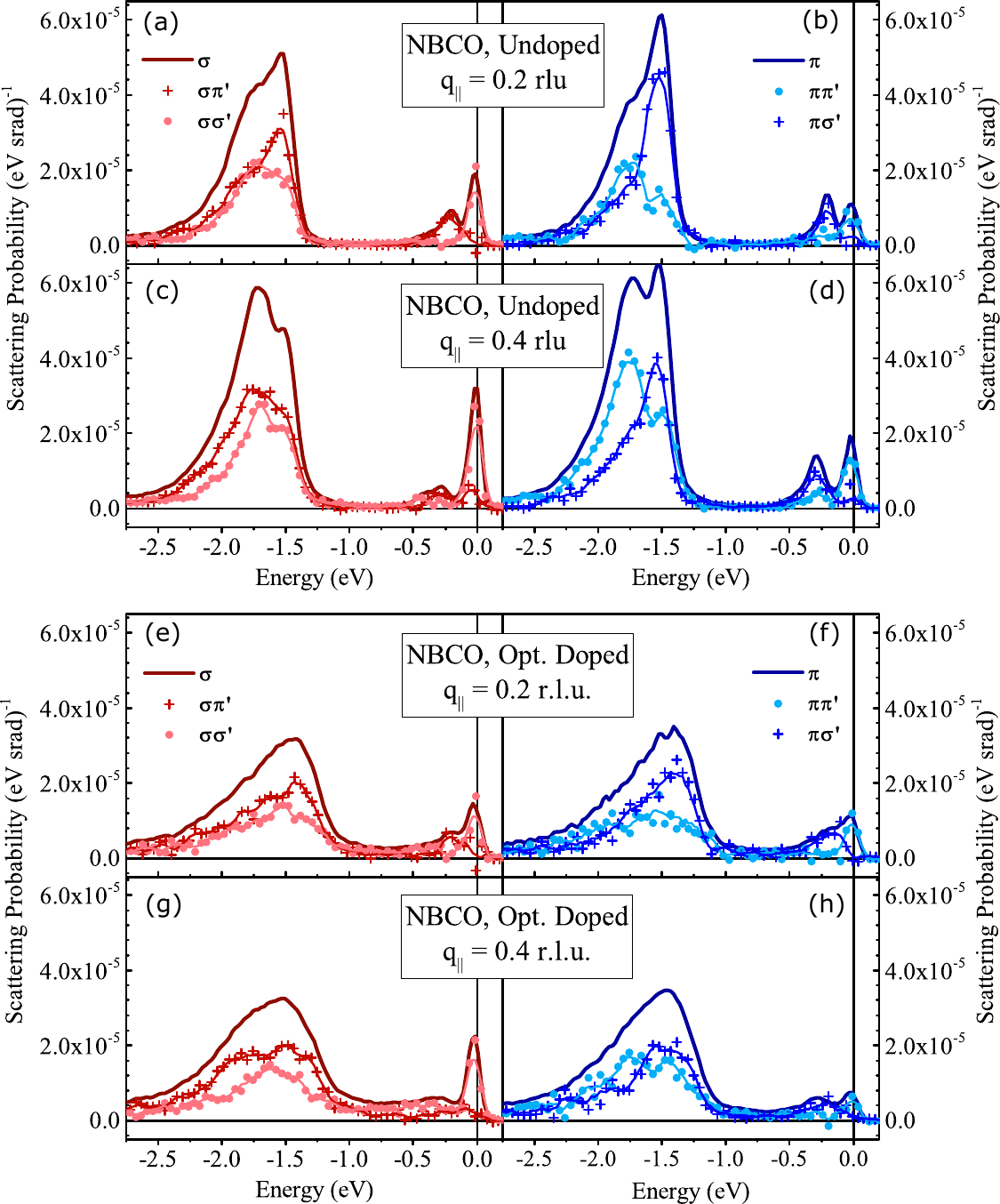}
	\caption{\label{fig:Panel} Polarization resolved RIXS spectra of AF (top panels) and OP (bottom panels) NBCO at $q_{\|}$ = 0.2 rlu and $q_{\|}$ = 0.4 rlu taken with $\sigma$ (panels a,c,e,g) and $\pi$ (panels b,d,f,h) polarization of the incident light. In the decomposed spectra symbols represent raw data while continuous lines are smoothed data on 7 points.}
\end{figure*}

\section{Results and Discussion}
\label{Results}

Fig.\,\ref{fig:Panel} shows some selected polarization-resolved RIXS spectra of AF and OP NBCO collected at $q_{\|}$ = 0.2 and 0.4 rlu, with both $\sigma$ and $\pi$ incident photon polarization. The intensity of crystal-field excitations shows both polarization and momentum transfer (scattering geometry) dependence, whereas their position does not.
At lower energy losses, in the mid-infrared region, the spectra exhibit a resolution-limited (quasi-)elastic line and dispersive inelastic features, related to magnetic excitations (magnon, bimagnon, multi-magnons). In the following, we will detail our analysis on each of these features.

\subsection{(Quasi-)elastic line and phonons}

\begin{figure*}[ht]
	\includegraphics[width=1.5\columnwidth]{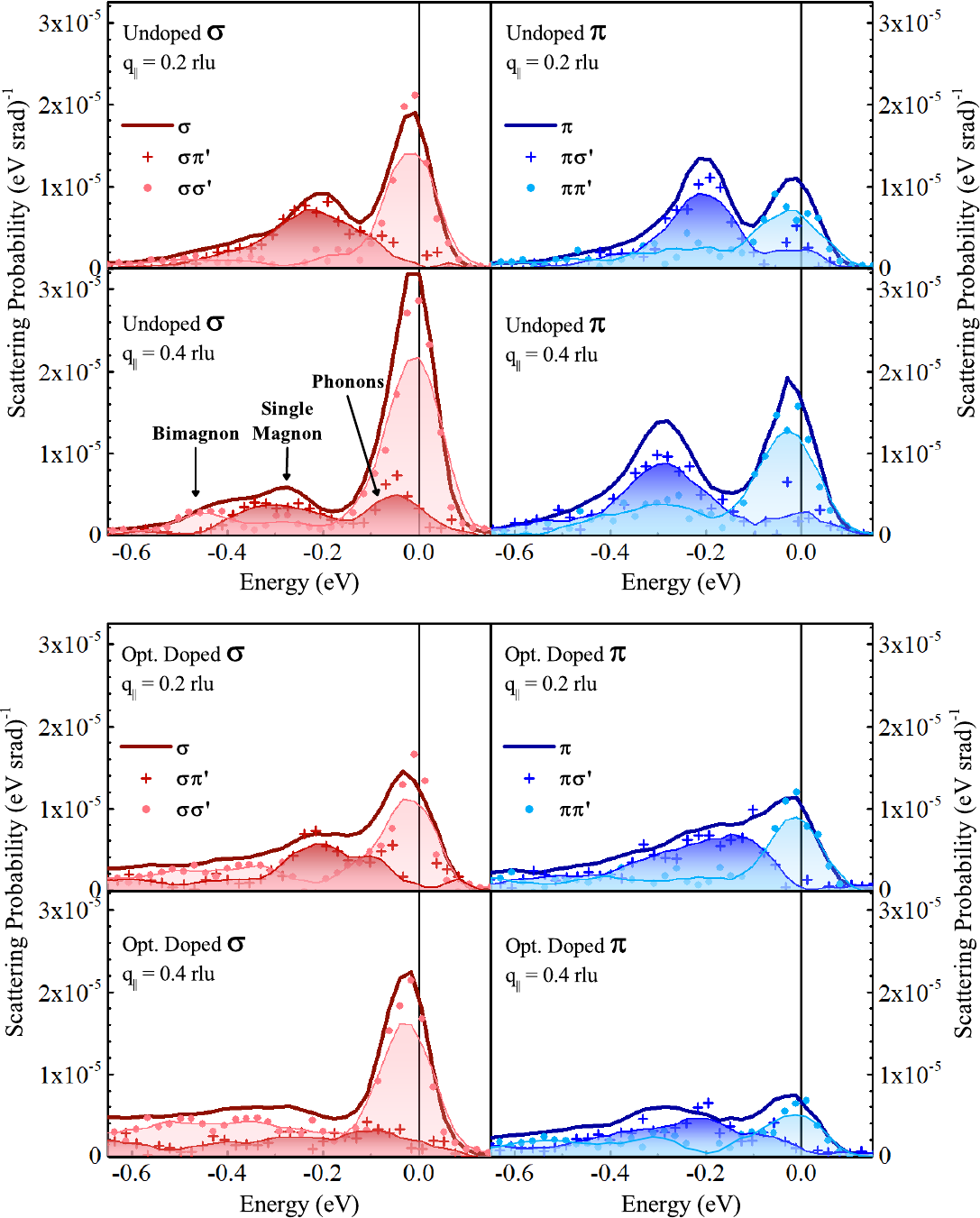}
	\caption{\label{fig:LowEnergy} Low energy region of the polarization-resolved RIXS spectra of AF (top panels) and OP (bottom panels) NBCO at $q_{\|}$ = 0.2 rlu and  $q_{\|}$ = 0.4 rlu. Shaded areas represent smoothed data on 7 points.}
\end{figure*}

Fig.\,\ref{fig:LowEnergy} focuses on the low-energy region of the RIXS spectra. In the case of AF NBCO (top panels), we notice that the (quasi-)elastic peak mainly belongs to the non-cross-polarization channels. Nevertheless, it retains some sizable contribution in the cross-polarization one. This observation apparently contradicts the predictions of the single-ion model, according to which (top panel of Fig.\,\ref{fig:ElasticSF}) spin-conserving ($\Delta S=0$) elastic scattering does not occur in the cross-polarization channels. This apparent inconsistency can simply be explained by the presence of low-energy excitations. In the case of cuprates, the two main phonon modes that are accessible by RIXS are the so-called buckling ($\approx$ 35 meV) and breathing ($\approx$ 70 meV) modes \cite{TomPRX2016}; the former is related to vibrations of the planar oxygen atoms in the direction perpendicular to CuO$_2$ planes, the latter to the Cu-O bond-stretching vibrations. While the experimental energy resolution of the present experiment is not sufficient for an accurate study of phonons, polarization analysis of the scattered photons provides clear evidence of their presence and possibly adds information about their symmetry. For example, we note that their occurrence in the cross-polarization channel is consistent with Raman measurements\cite{Thomsen2006}.
A detailed investigation of phonons, including the polarization analysis of their RIXS cross-section and dependence upon doping is out of the scope of this paper. However, here we would like to emphasize that the polarization analysis of the scattered photons, besides energy resolution, could provide crucial information about the symmetry of the phonon modes and eventually greatly contribute to uncover the nature of the electron-phonon coupling in undoped and superconducting cuprates.

\subsection{Magnetic Excitations}

\begin{figure}[h]
	\includegraphics[width=1\columnwidth]{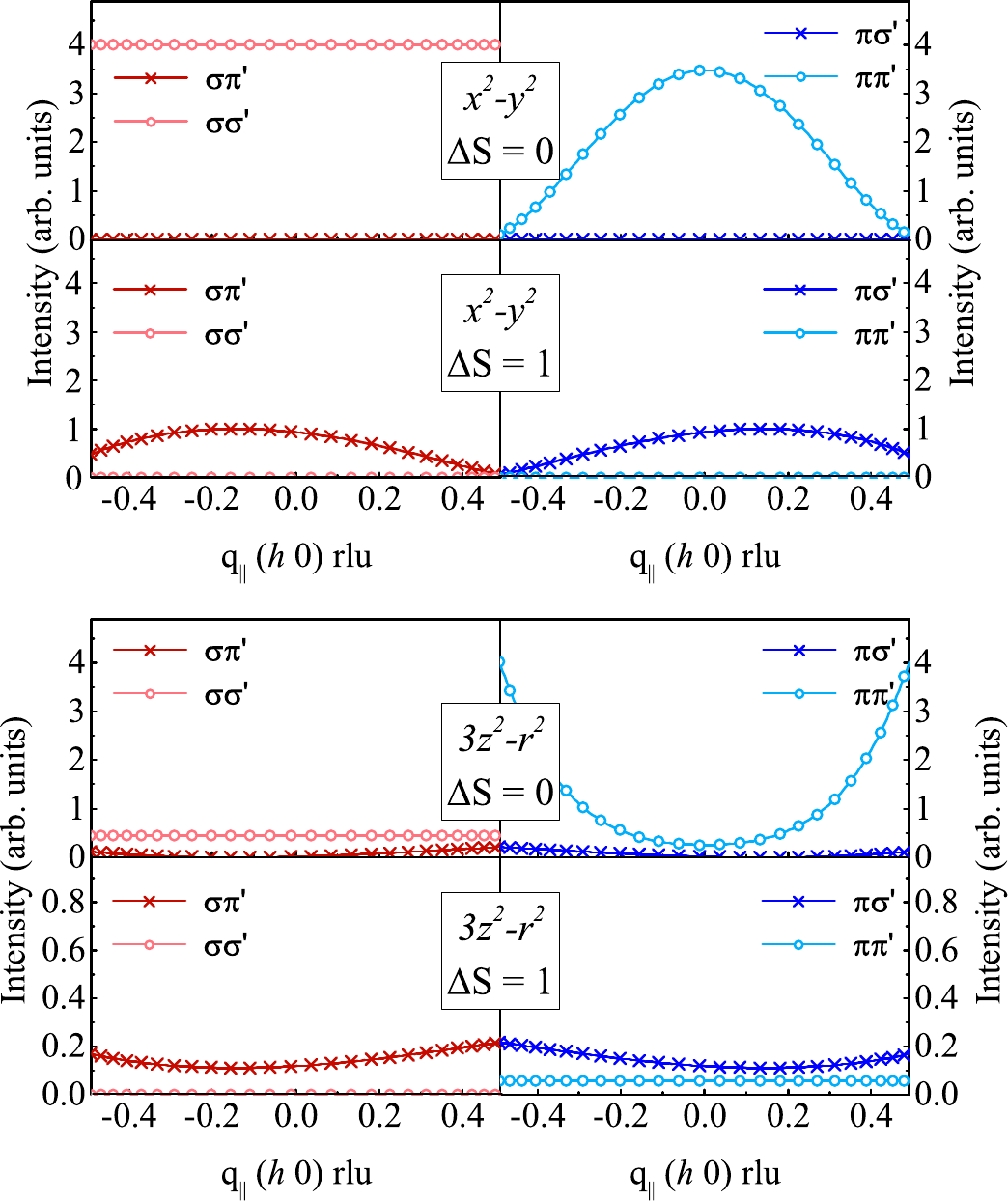}
	\caption{\label{fig:ElasticSF} In-plane momentum dependence of the polarization-resolved RIXS cross-sections within the single-ion model for excitations without orbital character. Top (bottom) panel assumes a ground state with pure $x^2-y^2$ ($3z^2-r^2$) symmetry. The scattering angle (2$\theta$) has been fixed to 149.5$^\circ$.}
\end{figure}

Starting from the case of AF NBCO (top panels of Fig.\,\ref{fig:LowEnergy}), we notice that magnons (the sharp peaks found at $\approx -250$ meV at 0.2 rlu and at $\approx -300$ meV at 0.4 rlu) occur predominantly in the cross-polarization channels. However, some sizeable contribution in the non-cross-polarization channels is visible. Again, this observation apparently contradicts the prediction of the single-ion model that (top panel of Fig.\,\ref{fig:ElasticSF}) spin-flip excitations ($\Delta S=1$) occur exclusively in the cross-polarization channels \cite{Ament2009}. We envisage two possible explanations for this discrepancy: i) the bimagnon continuum (see below) leaks spectral weight in the energy range of the spin-flip excitations; ii) the ground state of NBCO is not purely $x^2-y^2$, but it is mixed with the $3z^2-r^2$ state, a scenario that has been already considered before \cite{Sakakibara2010}. In order to check the consistency of the latter hypothesis, we calculated in the bottom panel of Fig.\,\ref{fig:ElasticSF} the corresponding RIXS cross-sections and verified that spin-flip transitions are allowed in the $\pi\pi^\prime$, but not in the $\sigma\sigma^\prime$ polarization channel. Although both effects could be simultaneously at play, the bimagnon contribution seems more likely to dominate. 

In the RIXS spectra measured with $\sigma$ polarized incident photons, a shoulder of the spin-flip excitations is usually ascribed to the bimagnon continuum, \emph{e.g.} the excitations of two interacting magnons. Unlike O $K$-edge, Cu $L_3$-edge RIXS probes a dispersive branch of the bimagnon continuum \cite{Bisogni2012BimagnonLedge} and therefore is more visible at large $q_{\|}$. At $q_{\|}=0.4$ rlu the bimagnon is found at an energy loss of approximately 450 meV, that is slightly above the single magnon peak. Contrary to single magnons, bimagnons mostly occur in the non-crossed-polarization $\sigma\sigma^{\prime}$ channel.

The bottom panels of Fig.\,\ref{fig:LowEnergy} show polarization-resolved RIXS spectra of optimally doped NBCO. Compared to the undoped sample, RIXS features in doped NBCO are broader and sit on an electron-hole pair excitation continuum, which mainly belongs to the non-cross polarization channel. We notice that magnons are heavily damped (which is why they are usually called paramagnons), but persist when mobile holes are added to the system, as it was already pointed out earlier \cite{LucioMagnonRIXSPRL2010,MLTparamagnonsNatPhys,DeanNatMat2013,DeanPRL2013,MLTPRB2013,YYPRB2015,MinolaPRL2015}. In addition, here we show that they preserve the polarization dependence of spin-flip excitations, as it is better evidenced in the case of $\pi$ incident photon polarization. 

\subsection{Crystal-field excitations}

In the previous section we discussed the polarization dependence of crystal-field or $dd$ excitations in AF NBCO. The absence of energy-momentum dispersion and the remarkable agreement with simulated RIXS spectra underline the localized (non-collective) nature of crystal-field excitations in NBCO. It is therefore interesting to study their evolution when charge carriers are introduced into the system. 

\begin{figure}[h]
	\includegraphics[width=1\columnwidth]{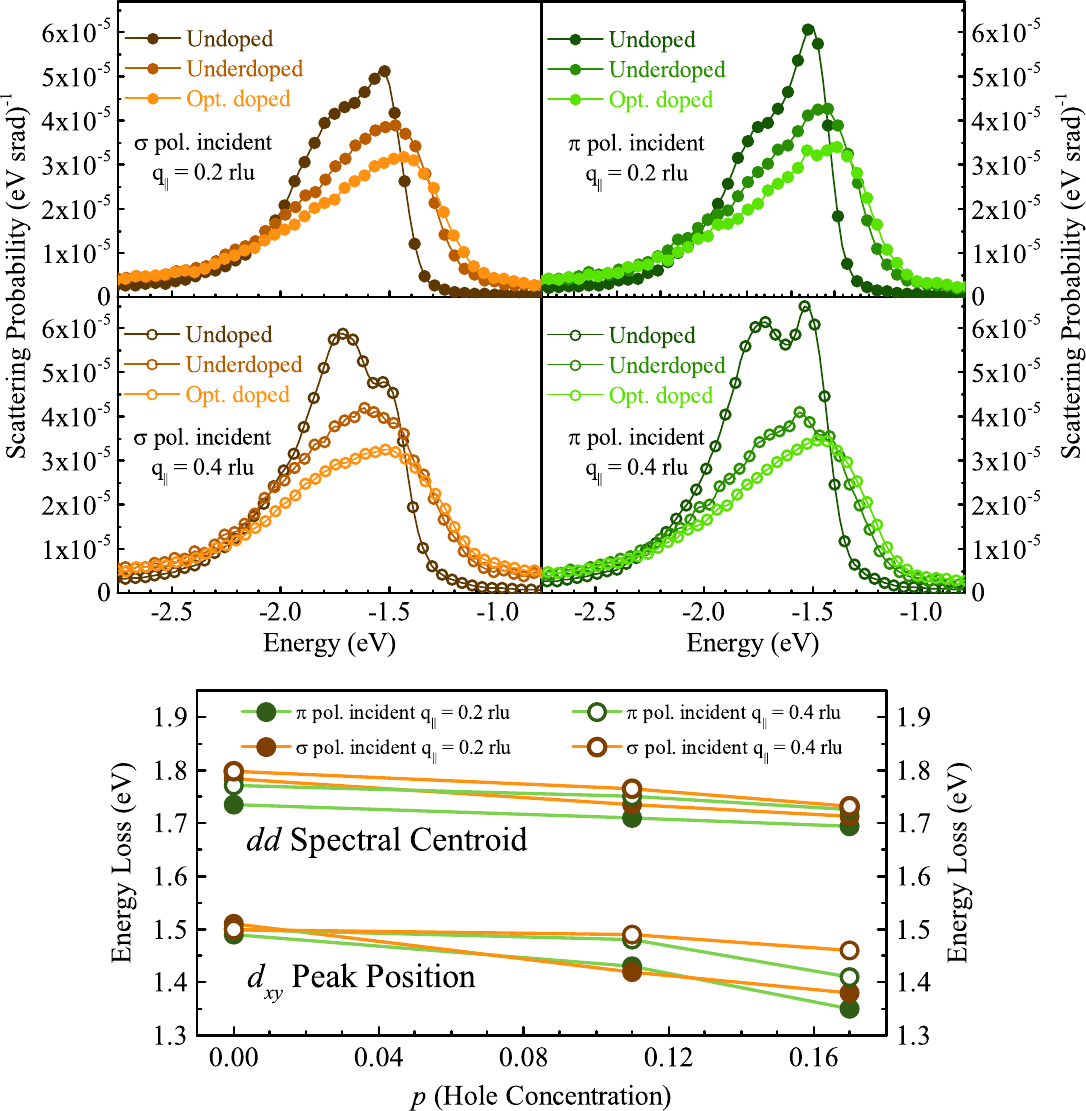}
	\caption{\label{fig:ddanalysis} Doping dependence of $dd$ excitations at $\textbf{q}_{\|}$ = 0.2 rlu and $\textbf{q}_{\|}$ = 0.4 rlu taken with $\pi$ and $\sigma$ incident light polarization. The bottom panel shows the $dd$ spectral centroids and the $xy$ peak positions.}
\end{figure}

In Fig.\,\ref{fig:ddanalysis} (top panels) we show the doping dependence (AF, UD and OP) of $dd$ excitations with no polarization analysis of the scattered photons for $q_{\|}$ = 0.2 and 0.4 rlu and for both $\sigma$ and $\pi$ incident photon polarizations. As the hole doping is increased we notice two major effects: i) the peaks corresponding to the different orbital excitations broaden and increasingly overlap with each other and ii) the energy position of the resulting broad distribution moves to lower energy loss. In order to quantify this latter effect we determined the center of mass of the distribution of $dd$ excitations and estimated the position of the lowest crystal field excitation ($xy$) by taking the second derivative of the spectra. We did not use the results of a multi-peak fitting procedure of the RIXS spectra of doped NBCO because it is affected by a very large uncertainty of the fitting parameters, but in the case of AF NBCO the energy of the transition to the ${xy}$ state obtained with the second derivative method perfectly agrees with the value (-1.52 eV) from Ref.\,\onlinecite{ddMoretti}. The analysis is summarized in the bottom panel of Fig.\,\ref{fig:ddanalysis}: the average softening of $dd$ excitations amounts to approximately 50 meV in the explored doping range and seems to be mostly caused by a substantial shift (150 meV) of the ${xy}$ state to lower energy losses. 

\begin{figure}[h]
	\includegraphics[width=1\columnwidth]{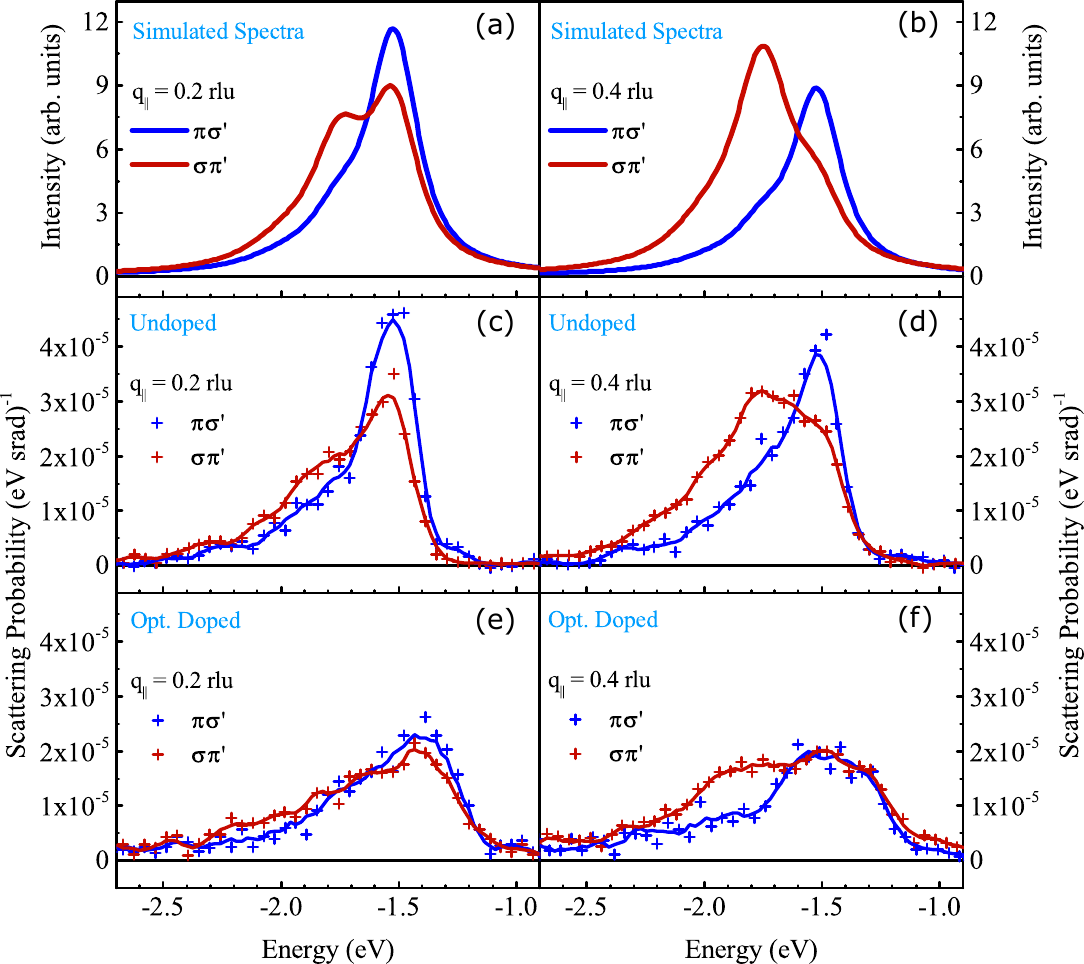}
	\caption{\label{fig:ddcrossed} (a-b) Simulated $\pi\sigma'$ and $\sigma\pi'$ spectra of AF NBCO for momentum transfers of $q_{\|}$ = 0.2 rlu and $q_{\|}$ = 0.4 rlu (orbital energies are taken from Ref.\,\onlinecite{ddMoretti}). Comparison between cross-channel polarization spectra of AF (c,d) and OP (e,f) NBCO at momentum transfers $q_{\|}$ = 0.2 rlu (left panels) and $q_{\|}$ = 0.4 rlu (right panels). Continuous lines are smoothed data on 7 points.}
\end{figure}

In Fig.\,\ref{fig:ddcrossed} we take advantage of the polarization resolution to further investigate this effect in AF and OP NBCO. We focus in particular on the RIXS spectra at $q_{\|}$ = 0.4 rlu in the cross-polarization channels: according to the calculated RIXS cross-sections, the $\pi\sigma^\prime$ polarization channel enhances the contribution of the $xy$ excited state and the $\sigma\pi^\prime$ polarization channel that of the $xz/yz$ states in this scattering geometry, implying that we could follow the evolution of the different states independently. The fact that the simulated RIXS spectra in panel (b) nicely reproduce the experimental data of AF NBCO in panel (d) supports the feasibility of this approach. Panel (f) shows that the main effect of doping on the $xy$ and $xz/yz$ excited states is broadening. On top of that, additional spectral weight on the low-energy side of the $xy$ state can be seen in the $\pi\sigma^\prime$ (blue) curve; on the contrary, the  $\sigma\pi^\prime$ (red) curve shows additional spectral weight on the high-energy side of the $xz/yz$ states. The effect is smaller and opposite than that for the $xy$ excited state, so that the center of mass of the whole $dd$ excitations distribution on average moves to lower energy losses. The RIXS spectra measured at $q_{\|}$ = 0.2 rlu (panels (c) and (e)) also provide evidence for the softening of the $xy$ excited state. The shift of the crystal-field states can be explained by the partial screening of the (negative) oxygen charges by doping holes thus reducing the effective crystalline electric field. That the effect is larger for in-plane ($xy$) than for out-of-plane ($xz/yz$) orbitals is consistent with the formation of Zhang-Rice singlets\cite{ZhangRice}, which mostly live in the CuO$_2$ planes, thereby providing a better screening of the in-plane oxygen charges.

\section{Conclusions}

In this work we provided a comprehensive polarization-resolved RIXS study of NBCO as a function of doping. We looked at a number of features, including phonons, single spin-flip excitations (magnons), bimagnons and $dd$ excitations and studied their full polarization dependence in RIXS. We confirmed that single-ion RIXS cross-section calculations reproduce most of the experimental observations, even including the polarization resolution of the scattered photons. By exploiting the differences in their RIXS cross-sections, we tracked the evolution of the various excitations as a function of doping. In particular, we reported a broadening and shift of the center of mass of the whole distribution of $dd$ excitations to lower energy losses. The analysis of the RIXS spectra in the cross-polarization channel allowed us to discriminate between $xy$ and $xz/yz$ excited states and confirm that the shift of $dd$ excitations is mainly driven by a softening of the $xy$ state.

Finally, we emphasize that precautions should be taken in order to obtain a correct interpretation of polarization-resolved RIXS spectra. This was discussed with the introduction of the Stokes (and Poincarè-Stokes) parameters: in particular, it turns out that the decompositions in $\sigma^\prime$ and $\pi^\prime$ components is rigorous only when the scattered radiation is fully $\sigma^\prime$ or $\pi^\prime$ polarized, namely $|S^\prime_1|=S^\prime_0$ ($|P^\prime_1|=1$). Single-ion model calculations show that this is the case for most of the excited states. In general, we believe that a systematic use of polarization-resolved RIXS could add important information on the nature of the excitations; for example, it may help to discriminate phonon modes with different symmetries. We note that only the knowledge of the polarization of the scattered photons permits a proper correction for self-absorption effects. This might be crucial when investigating little intensity differences, especially in the low-energy region of the RIXS spectra where phonons are observed. In addition, polarization resolution of the scattered photons mitigates issues related to the intrinsic broadening of the RIXS features upon doping, for which pushing energy resolution does not necessarily help, \emph{e.g.} for magnons, bimagnons and electron-hole pair excitations overlapping in the same energy range or for crystal-field excitations, which broaden and merge into a single feature.

\section*{Acknowledgements}
\label{Acknowledgements}

The experimental data were collected at the beam line ID32 of the European Synchrotron (ESRF) in Grenoble (France) using the ERIXS spectrometer designed jointly by the ESRF and the Politecnico di Milano. This work was supported by ERC-P-ReXS project (2016-0790) of the Fondazione CARIPLO, Regione Lombardia and by MIUR Italian Ministry for Research through project PIK Polarix. M.M. and H. S. were partially supported by the Alexander von Humboldt Foundation.

\section*{Appendix}
\appendix

\section{Poincar\'e sphere}
\label{PoincareStokes}

\begin{figure}[hb]
	\includegraphics[width=1\columnwidth]{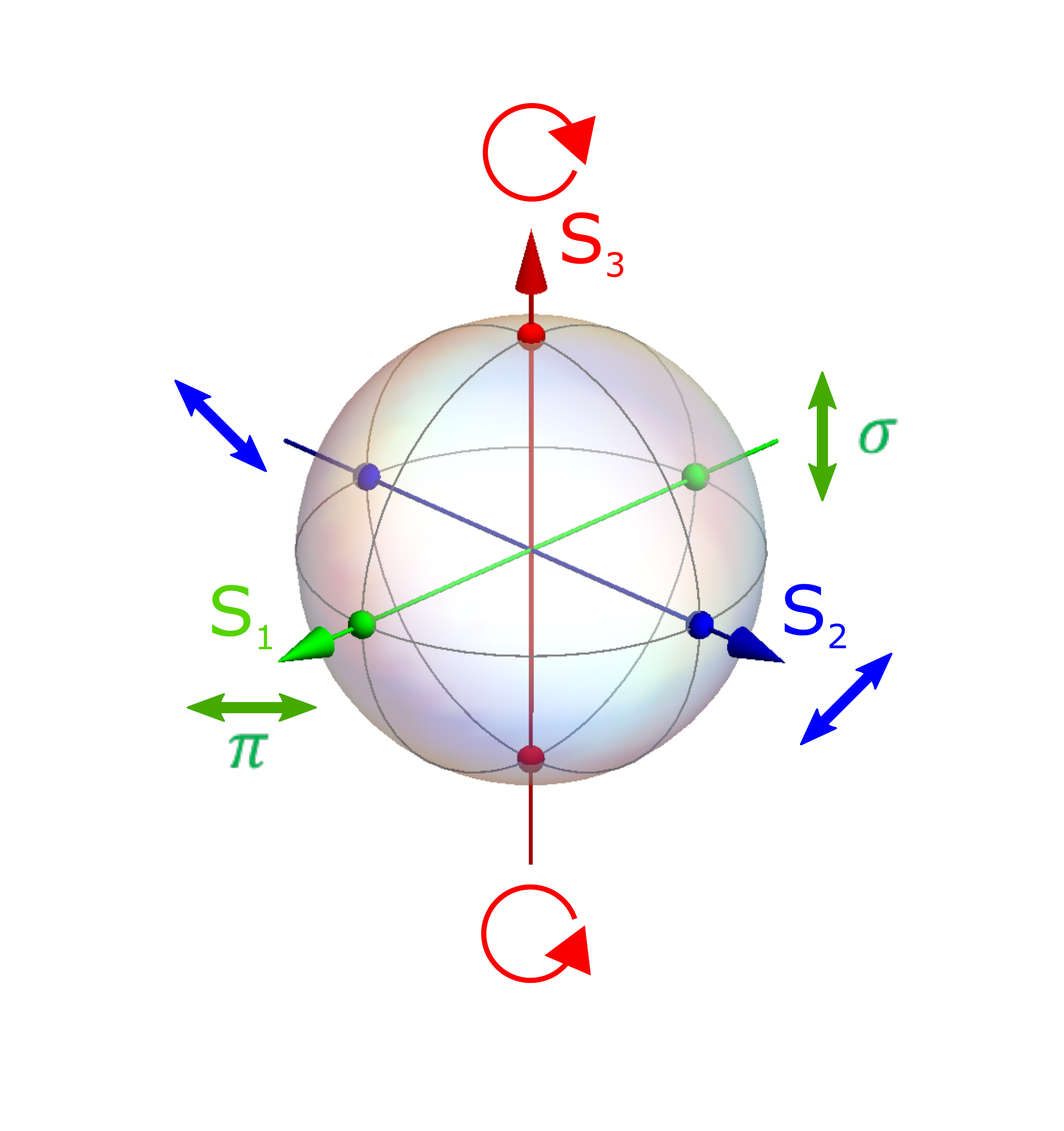}
	\caption{\label{fig:Ellipse}  Schematic representation of the Poincar\'e sphere. The Cartesian coordinates represent the three Stokes parameters $S_1$, $S_2$ and $S_3$.}
\end{figure}

Fully polarized electromagnetic radiation is described by the (complex) components of the electric field in the coordinate axes, say $\bm{\epsilon}_\sigma$ and $\bm{\epsilon}_\pi$, perpendicular to the propagation direction $\mathbf{k}$. These can be regarded as the components of a vector 
\begin{equation}
\mathbf{E}=
\begin{bmatrix}
E_\sigma \\
E_\pi
\end{bmatrix}=
\begin{bmatrix}
|E_\sigma|e^{i \delta_\sigma} \\
|E_\pi|e^{i \delta_\pi}
\end{bmatrix}
\end{equation}
known as the Jones vector. The action of any optical element on the properties of the electromagnetic field is described by the so-called Jones matrix $\mathcal{J}$. For reflection, it reads
\begin{equation}
\mathcal{J}_r=\begin{bmatrix}
-\sqrt{r_{\sigma}} & 0\\
0 & \sqrt{r_{\pi}}
\end{bmatrix}.
\end{equation}
and the Jones vector of the reflected radiation is therefore given by
\begin{equation}
\mathbf{E}_r=\mathcal{J}_r \mathbf{E}= 
\begin{bmatrix}
-\sqrt{r_{\sigma}} E_\sigma \\
\sqrt{r_{\pi}}E_\pi
\end{bmatrix}=
\begin{bmatrix}
-\sqrt{r_{\sigma}}|E_\sigma|e^{i \delta_\sigma} \\
\sqrt{r_{\pi}}|E_\pi|e^{i \delta_\pi}
\end{bmatrix}.
\end{equation}

Alternatively, the polarization state of the radiation can be described by the four Stokes parameters, forming the Stokes vector $\mathbf{S}$. The relation to the components of the Jones vector can be written in several, equivalent ways

	\begin{eqnarray}
	\mathbf{S} &=&\begin{bmatrix}
	S_{0}\\
	S_{1}\\
	S_{2}\\
	S_{3}
	\end{bmatrix}=\begin{bmatrix}
	|E_{\sigma}|^2+|E_{\pi}|^2\\
	|E_{\sigma}|^2-|E_{\pi}|^2\\
	\frac{1}{2}\left(|E_{\sigma}\!+\!E_{\pi}|^2-|E_{\sigma}\!-\!E_{\pi}|^2 \right)\\
	\frac{1}{2}\left(|E_{\sigma}\!-\!\imath E_{\pi}|^2-|E_{\sigma}\!+\!\imath E_{\pi}|^2 \right)
	\end{bmatrix}= \nonumber \\
	&=& \begin{bmatrix}
	E_{\sigma}E_{\sigma}^\ast + E_{\pi} E_{\pi}^\ast\\
	E_{\sigma}E_{\sigma}^\ast - E_{\pi} E_{\pi}^\ast\\
	E_{\sigma}E_{\pi}^\ast + E_{\sigma}^\ast E_{\pi}\\
	\imath (E_{\sigma}E_{\pi}^\ast - E_{\sigma}^\ast E_{\pi})\end{bmatrix}=\begin{bmatrix}
	E_{\sigma}E_{\sigma}^\ast + E_{\pi} E_{\pi}^\ast\\
	E_{\sigma}E_{\sigma}^\ast - E_{\pi} E_{\pi}^\ast\\
	2\Re\{E_{\sigma}E_{\pi}^\ast\}\\
	2\Im\{E_{\sigma}^\ast E_{\pi}\}
	\end{bmatrix}= \nonumber \\
    &=& \begin{bmatrix}
	|E_{\sigma}|^2+|E_{\pi}|^2\\
	|E_{\sigma}|^2-|E_{\pi}|^2\\
	2|E_{\sigma}| |E_{\pi}| \cos(\delta_\sigma-\delta_\pi)\\
	-2|E_{\sigma}| |E_{\pi}| \sin(\delta_\sigma-\delta_\pi)
	\end{bmatrix}.
	\end{eqnarray} 
$S_{0}$ is the total intensity of the electromagnetic radiation, $S_{1}$ and $S_{2}$ describe the degree of linear polarization and $S_{3}$ that of circular polarization. In addition, one can define the Poincaré-Stokes parameters $\mathbf{P}$, where $P_{i}=S_{i}/S_0$ ($i=1,2,3$). In general, $S_1^2 + S_2^2 + S_3^2 \leq S_0^2$ ($P_1^2 + P_2^2 + P_3^2 \leq 1$), where the equality holds for fully polarized radiation, in which case the Poincaré-Stokes parameters defines a point on the so-called Poincaré sphere of unit radius, as shown in Fig.\,\ref{fig:Ellipse} .

The propagation of electromagnetic radiation characterized by $\mathbf{S}$ through an optical element is described by the M\"{u}ller matrix $\mathcal{M}$, which is related to the elements of $\mathcal{J}$\cite{}, by $\mathcal{M} = A(\mathcal{J}^\ast\otimes\mathcal{J})A^{-1}$, where
\begin{equation}
A=\begin{bmatrix}
1 & 0 & 0 & 1 \\
1 & 0 & 0 & -1 \\
0 & 1 & 1 & 0 \\
0 & \imath & -\imath & 0
\end{bmatrix}.
\end{equation}
For reflection optics, the M\"{u}ller matrix reads
\begin{equation}
\mathcal{M}_r =\begin{bmatrix}
\frac{1}{2}(r_{\sigma}+r_{\pi}) & \frac{1}{2}(r_{\sigma}-r_{\pi}) & 0 & 0 \\
\frac{1}{2}(r_{\sigma}-r_{\pi}) & \frac{1}{2}(r_{\sigma}+r_{\pi}) & 0 & 0 \\
0 & 0 & \sqrt{r_{\sigma} r_{\pi}} & 0 \\
0 & 0 & 0 & \sqrt{r_{\sigma} r_{\pi}}
\end{bmatrix}
\end{equation}
and its action on the Stokes vector $\mathbf{S}$ determines how the properties of the electromagnetic radiation change after reflection
\begin{equation}
\mathbf{S}_r=\mathcal{M}_r \mathbf{S}
=\begin{bmatrix}
S_{r,0}\\
S_{r,1}\\
S_{r,2}\\
S_{r,3}
\end{bmatrix}=\begin{bmatrix}
\frac{1}{2}\left[r_{\sigma}(S_0+S_1) + r_{\pi}(S_0-S_1)\right] \\
\frac{1}{2}\left[r_{\sigma}(S_0+S_1) - r_{\pi}(S_0-S_1)\right] \\
\sqrt{r_{\sigma} r_{\pi}}S_2\\
\sqrt{r_{\sigma} r_{\pi}}S_3
\end{bmatrix}.
\end{equation} 
Again, $S_{r,0}$ is the total intensity of the reflected electromagnetic radiation and similarly for the other parameters.

\section{Error bars}
\label{Errorbars}

In the following, we report the procedure we adopted to calculate error bars associated with the intensity of the RIXS spectra with polarization analysis of the scattered photon polarization. Since the polarization-resolved RIXS spectra are derived indirectly from two independent measurements of the RIXS intensities in the ``direct'' beam ($I$) and past the multilayer mirror ($I_M$), the error propagation is non-trivial. In the following, we calculate error bars with the same approach used for spin-resolved photoemission with Mott detectors \cite{Kessler1985PolarizedElectrons}.

The two RIXS spectra $I$ and $I_M$ are typically measured with different acquisition times ($\tau \neq \tau_M$) in order to compensate for the low efficiency of the multilayer mirror. The total numbers $I$ and $I_M$ of incident photons on the two detectors are:

\begin{eqnarray}
I &=& (n_{\pi^{\prime}}+n_{\sigma^{\prime}})\tau=n \tau \\ 
I_M &=&(r_{\pi
}n_{\pi^{\prime}}+r_{\sigma^\prime}n_{\sigma^{\prime}})\tau_M = \nonumber \\ 
&=& n_M \tau_M,
\label{equ:errorbars1}
\end{eqnarray}

\noindent where $r_{\pi^\prime}$ ($r_{\sigma^\prime}$) is the multilayer mirror reflectivity for the $\pi$ ($\sigma$) polarization channel and $n_{\pi^\prime}$ ($n_{\sigma^\prime}$) is the energy-dependent number of $\pi^\prime$ ($\sigma^\prime$) polarized scattered photons hitting the detector per unit time.

Having introduced the Sherman constant for the polarimeter $A$, the reflectivity of the multilayer mirror can be written as $r_{\sigma,\pi} = r_0 (1 \mp A)$ (see the main text for the definitions of $A$ and $r_0$). In order to proceed with the calculation of the error bars of the polarization-resolved RIXS intensities, we define the degree of photon polarization
\begin{equation}
P = \frac{n_{\sigma^{\prime}}-n_{\pi^{\prime}}}{n_{\sigma^{\prime}}+n_{\pi^{\prime}}}
\end{equation}
and the polarization state of the scattered photons by the sample
\begin{equation}
D =  AP = \frac{I_M \tau}{I \tau_M r_0} -1.
\label{equ:errorbars2}
\end{equation}

\noindent Considering that the total number of detected photons is given by $I + I_M$ and assuming a Poisson statistical distribution of uncertainties, the error bar for both polarization channels is given by
\begin{eqnarray}
\Delta D &=&  \Bigg[\left(\frac{\partial D}{\partial I}\right)^{2} \Big(\Delta I\Big)^{2} + \nonumber \\
&+& \left(\frac{\partial D}{\partial I_M}\right)^{2} \Big(\Delta I_M\Big)^{2} \Bigg]^{1/2} = \nonumber \\
&=& \frac{\tau}{\tau_M r_0}\sqrt{\frac{{I_M}^{2}+I_M I}{{I}^{3}}}.
\label{equ:errorbars3}
\end{eqnarray}
where $\Delta I$ ($\Delta I_M$) is the error bar on the intensity of the direct beam (the beam past the multilayer). Now, rewriting $n_{\pi^\prime}$ and $n_{\sigma^\prime}(\omega_2)$ as
\begin{equation}
n_{\pi^{\prime},\sigma^{\prime}}(\omega_2) = (1 \mp P)\frac{I}{2\tau} 
\end{equation}
and considering that $\Delta P  = \Delta D /A$, Eq.\,(\ref{equ:errorbars3}) can be casted in the form
\begin{equation}
\Delta n_{\pi^{\prime},\sigma^{\prime}}  =\frac{1}{2A\tau_M r_0}\sqrt{\frac{{I_M}^{2}+I_M I}{I }},
\label{equ:errorbars4}
\end{equation}
\noindent or, alternatively, 
\begin{eqnarray}
\Delta I_{\pi^{\prime},\sigma^{\prime}} &=&\tau \Delta  n_{\pi^{\prime},\sigma^{\prime}} =  \nonumber \\ 
&=& \frac{\tau}{2A\tau_M r_0}\sqrt{\frac{{I_M}^{2}+I_M I}{I}}.
\label{equ:errorbars5}
\end{eqnarray}

\noindent From Eq.\,\ref{equ:errorbars5} we notice that if the total number of accumulated counts is roughly the same for the direct beam and the beam past the multilayer, we have $I \simeq I_M$, implying that $\tau_M \simeq \tau/r_0$ and $\Delta I_{\pi^{\prime},\sigma^{\prime}} \simeq \sqrt{n(\omega_2)}/(A\sqrt{2})$.

\bibliography{BibliographyPaperPolarimeter}

\end{document}